\documentclass[nofootinbib,showpacs,showkeys]{revtex4-2}
\usepackage{graphicx,color}
\usepackage{amssymb}
\usepackage{amsmath}
\usepackage{bm}
\usepackage{tensor} 
\usepackage{lipsum}
\usepackage{latexsym}
\usepackage{dcolumn}
\usepackage{float}
\usepackage{cancel}
\usepackage[hyperindex,colorlinks]{hyperref}

 \usepackage{cleveref}
\usepackage{graphicx}
\usepackage{dcolumn}
\usepackage{bm}
\usepackage{tensor} 
\usepackage{natbib}
\usepackage{subcaption}
\graphicspath{{Images/}}

\makeatletter
\newcommand{\addbar}[3]{{\vphantom{#3}\mathpalette\add@bar{{#1}{#2}{#3}}}}
\newcommand{\add@bar}[2]{\add@@bar{#1}#2}
\newcommand{\add@@bar}[4]{%
  
  \begingroup
  \sbox\z@{$\m@th#1#4$}%
  \ooalign{%
    \hidewidth\kern#2\wd\z@\add@@@bar{#1}{#3}\hidewidth\cr
    \box\z@\cr
  }%
  \endgroup
}
\newcommand{\add@@@bar}[2]{%
  \sbox\tw@{$\m@th#1\newmcodes@\if\relax#2\relax-\else\bm{-}\fi$}%
  \raisebox{\dimexpr(\ht\z@-\ht\tw@)/2}{\usebox\tw@}%
}
\makeatother

\usepackage{acro}

\DeclareAcronym{kds}{
  short=KdS,
  long=Kerr-de Sitter,
}
\DeclareAcronym{rkds}{
  short=RKdS,
  long=Kerr-de Sitter Revisited,
}
\DeclareAcronym{sds}{
  short=SdS,
  long=Schwarzschild-de Sitter,
}
\DeclareAcronym{eht}{
  short=EHT,
  long=Event Horizon Telescope,
}
\DeclareAcronym{ligo}{
short=LIGO,
long=Laser Interferometer Gravitational-Wave Observatory
}
\begin{document}

\title{Remarks on the black hole shadows in Kerr-de Sitter space times}

\author{Eunice Omwoyo}%
\email{eunice.m.omwoyo@aims-senegal.org}
\affiliation{%
PPGCosmo, CCE, Universidade Federal do Esp\'irito Santo (UFES)\\
Av. Fernando Ferrari, 540, CEP 29.075-910, Vit\'oria, ES, Brazil.}%

\author{Humberto Belich}%
\email{humberto.belich@ufes.br}
\affiliation{%
N\'ucleo Cosmo-ufes \& Departamento de F\'isica,  Universidade Federal do Esp\'irito Santo (UFES)\\
Av. Fernando Ferrari, 540, CEP 29.075-910, Vit\'oria, ES, Brazil.}

\author{J\'ulio C. Fabris}
\email{julio.fabris@cosmo-ufes.org}%
\affiliation{%
N\'ucleo Cosmo-ufes \& Departamento de F\'isica,  Universidade Federal do Esp\'irito Santo (UFES)\\
Av. Fernando Ferrari, 540, CEP 29.075-910, Vit\'oria, ES, Brazil.}%
\affiliation{%
National Research Nuclear University MEPhI, Kashirskoe sh. 31, Moscow 115409, Russia}%

\author{Hermano Velten}%
\email{hermano.velten@ufop.edu.br}
\affiliation{%
Departamento de F\'isica, Universidade Federal de Ouro Preto (UFOP), Campus Universit\'ario Morro do Cruzeiro, 35.400-000, Ouro Preto, Brazil}%
\date{\today}

\begin{abstract}

This work is geared towards analysis of shadows cast by  \ac{kds} and \ac{rkds} black holes. Considering observers in the vicinity of the static radius, we derive the impact parameters defining the apparent positions of the shadows. Such observers are of interest to our work because embedding diagrams have shown that de Sitter space-time is analogous to an asymptotically flat one in the vicinity of the static radius. We also perform a comparative analysis between our result with that in Ref.\cite{ovalle2021kerr}. Furthermore, we numerically obtain the radii of curvature, vertical diameters and horizontal diameters of the shadows. We find that for $\Lambda=1.11\times 10^{-52} m^{-2}$, M87* observations cannot distinguish a \ac{rkds} black hole shadow from that of a Kerr black hole. Additionally, for the same value of $\Lambda$, \ac{kds} and \ac{rkds} black hole shadows are, in practise,  indistinguishable. Previously, it has also been shown that when $\Lambda=1.11\times 10^{-52} m^{-2}$,  \ac{kds} and Kerr black hole shadows are indistinguishable. Utilizing the 2017 EHT observations of M87* on the allowed range of the characteristic radius of the shadow, we obtain constraints on both black holes. When, $a/M>0.812311$, we observe that large angles of inclination ($\theta>30.5107^{\circ}$) do not pass the constraints for both \ac{kds} and \ac{rkds} black holes.
\end{abstract}

\maketitle

\section{Introduction}
In the General Theory of Relativity, the gravitational force is seen as the effect of the deformation of space-time. Black holes are regions of very strong gravity that are sufficient to warp space, bend light and give rise to space-time singularities. Since black holes results from the supernovae explosions that leave behind a positive angular momentum of the remaining matter, it is most probable that all black holes in nature are rotating and are therefore described by the Kerr solution \cite{kerr1963gravitational}. 

In the presence of a cosmological constant $\Lambda>0$, an expected component of the universe due to the dark energy phenomena, a generalization of the Kerr metric is given by the  \ac{kds} metric, a solution that was firstly found by Carter \cite{novikov1973houches}. The \ac{kds} metric belongs to the Plebanski-Demianski family of solutions \cite{debever1971type}. 
These solutions describe the most general stationary, axially symmetric Petrov Type-D metrics of Einstein-Maxwell equations with a cosmological constant. The Plebanski-Demianski solutions are characterized by seven parameters that in certain instances are related to acceleration, magnetic and electric charges, mass, cosmological constant, NUT parameter and angular momentum, \cite{kagramanova2008charged}. 

Recently, Ref. \cite{ovalle2021kerr} , through gravitational decoupling for axially symmetric systems (\cite{contreras2021gravitational},\cite{ovalle2017decoupling},\cite{ovalle2019decoupling}), have proposed a revisited solution, the  \ac{rkds} metric. \ac{rkds} solution is a rotating version of the  \ac{sds} solution, representing a black hole with a cosmological constant. The solution is asymptotically de Sitter and reduces to Kerr solution as a special case. Unlike the \ac{kds} solution, the \ac{rkds} solution is neither a $\Lambda-$Vacuum solution - it does not belong to the Plebanski-Demianski class of metrics - nor is it a constant curvature solution (i.e it exhibits warped curvature except on the equatorial plane where curvature remains constant). 

As a consequence of the cosmological constant, there arises a  cosmological horizon behind which the geometry of space-time is dynamic. For astrophysical processes, another radius associated with cosmic repulsion is relevant \cite{slany2020equatorial},
the so-called static radius. On the static radius boundary, gravitational attraction due to the central compact object and cosmic repulsion counterbalance each other. This radius  represents a natural boundary for gravitationally bound systems in an expanding universe \cite{stuchlik2020influence}. A cosmological horizon is relevant for cosmology and not for astrophysical processes that are limited by the static radius \cite{stuchlik2020influence}. 

When a black hole is in front of a luminous background, its unstable photon region, a region containing null geodesics at a constant radius, will be projected on the observer's sky to form the  so-called black hole shadow. The shadow appears as a dark disc with a bright ring around it. Static black holes cast circular shadows owing to their spherical symmetry. Rotating black holes on the other hand cast elongated shadows as a consequence of frame dragging effects. The field of studying black hole shadows as a feasible and reliable observational probe is still in the first days of life but the direct detection of gravitational waves by \ac{ligo} and Virgo collaboration \cite{abbott2016observation} and the images of the super-massive black hole at the centre of Messier 87 galaxy by the Event Horizon Telescope (EHT) \cite{event2019first}-\cite{akiyama2019first} have been advances of great significance towards observation of black holes.  

In this work, we investigate shadows cast by black holes in \ac{kds} and \ac{rkds} space-times. Our aim is to promote a comparative analysis between various results presented in the literature. The motivation for our study is as follows. In space-times that are non asymptotically flat, the calculation of the angular radius of the shadow solely requires the use of finite distance observers. For such observers, a choice of an orthonormal tetrad is necessary, as shown by \cite{li2020shadow}, \cite{grenzebach2014photon}  \cite{stuchlik2018light} and \cite{chowdhuri2021shadow}. In asymptotically flat space-times on the other hand, the angular radius of the shadow can be analogous to that in Minkowski space-time \cite{atamurotov2013shadow}. As a result, in asymptotically flat space-times, one calculates the parameters associated to the black hole shadows via the expressions
\begin{align}
\alpha= \lim\limits_{r_{o} \longrightarrow \infty}\left(-r_{o}^{2} \sin \theta_{o} \dfrac{d\phi}{dr}  \right), \label{al} \\
\beta=\lim\limits_{r_{o}\longrightarrow \infty}\left( r_{o}^{2}\dfrac{d\theta}{dr}  \right) \label{beta},
\end{align} 
\begin{figure}[htb]
	\centering 
	\begin{subfigure}{0.35\textwidth}
		\includegraphics[width=1.0\linewidth]{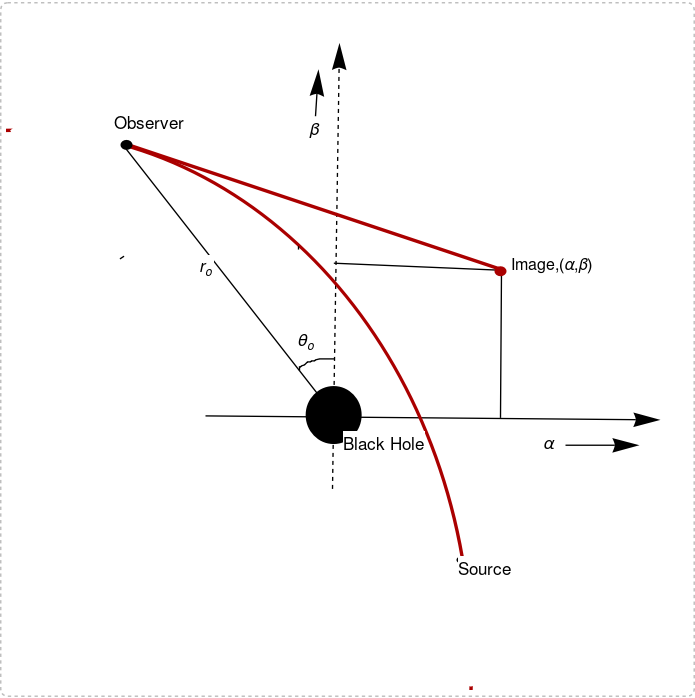}
		\caption{Illustration of the apparent positions ($\alpha$,$\beta$) of the image. The dotted line is the projection of the axis of symmetry on the celestial sphere. 
		}
		\label{fig:celc}
	\end{subfigure}\hfil
	\begin{subfigure}{0.35\textwidth}
		\includegraphics[width=1.0\linewidth]{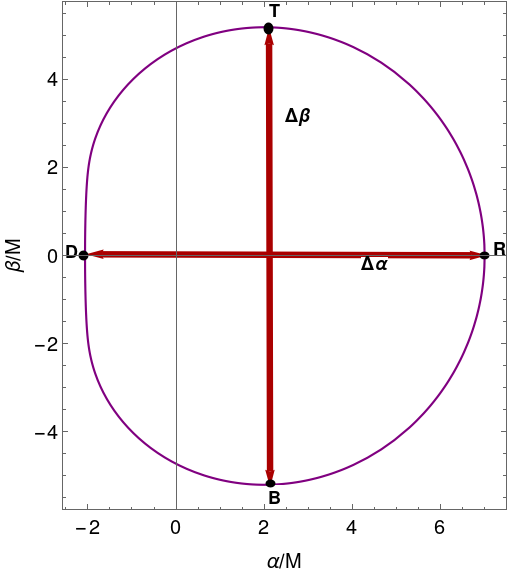}
		\caption{General contour for the image of a black hole. $T$,$D$,$R$,$B$,$\Delta \alpha$ and $\Delta \beta$ will be discussed in section V.}
		\label{fig:characpoints}
	\end{subfigure}\hfil
	\caption{ Illustration of the parameters $\alpha$ and $\beta$.
	}
	\label{fig:celcs}
\end{figure}
where $\alpha$ and $\beta$ are the impact parameters defining apparent positions of the image of the black hole on the celestial sphere \cite{cunningham1973optical}. Actually $\alpha$ is the apparent displacement of the image perpendicular to the projected axis of symmetry of the black hole while $\beta$ is the apparent displacement of the image parallel to the projected axis of symmetry \cite{cunningham1973optical}. $(r_{o},\theta_{o})$ are the positions of the observer. $\dfrac{d\phi}{dr}$ and $\dfrac{d\theta}{dr}$ are calculated from the null geodesic equations for the respective space-time. Properties of the geometry around black holes can be visualized by embedding their $2D$ sections of $t=$constant hypersurfaces onto  $3D$  Euclidean geometry. The resulting diagrams are known as embedding diagrams. Embedding diagrams expedite in acquiring an intuitive understanding of the gravitational field rendered into curved space-time. Besides, these diagrams aid in analyzing the effect of parameters such as magnetic field, electric charge, black hole spin or cosmological constant \cite{hledik2006visualizing}.  By the use of embedding diagrams of both ordinary geometry and  optical reference geometry, Refs. \cite{stuchlik1999some} and \cite{kovavr2006optical} have shown that in the vicinity of the static radius, the geometry of de Sitter space-time is analogous to an asymptotically flat space-time. Thus, \cref{al} and (\ref{beta}) can therefore be applied in the vicinity of the static radius where the \ac{kds} (or \ac{sds}) space-time is close enough to the asymptotically flat space-time case. This approach has been considered in Ref. \cite{ovalle2021kerr}, where the size of a shadow in \ac{kds} and \ac{rkds} space-times has been compared. We however note that there is a discrepancy on the analytical solution for the form of $\alpha$ and $\beta$ they use with the results we have obtained in our work. In Ref.\cite{ovalle2021kerr}, attention was not given to the analysis of the black hole shadows; they were proposing the RKdS solution for the first time and basically used the shadow to support their result that a KdS black hole has a larger event horizon than the RKdS black hole. As a result, the error does not affect their result. To our knowledge, there has been no work in the literature that has given a detailed analysis of the KdS and RKdS black hole shadows for observers in the vicinity of the static radius. Moreover, RKdS solution has just been recently proposed and hence has no detailed study of it's black hole shadows alongside a comparison with those of KdS. For astrophysical purposes, a detailed analysis of the black hole shadows is necessary and this is what we intend to do in this work. In our work by considering different black hole parameters, we have shown that the behaviour of the black hole shadows, for observers in the vicinity of the static radius, is consistent with the behaviour of the corresponding spherical photon orbits in both KdS and RKdS space-time.  We hope that from this work, more aspects of research can be considered in the vicinity of the static radius because a lot of study in KdS space-time has only been geared towards this space-time being asymptotically de sitter.
It is worth noting that black hole shadows can also be studied through other techniques such as general-relativistic radiative transfer and general-relativistic magnetohydrodynamic simulations (\cite{mizuno2018current},\cite{dexter2009millimeter},\cite{moscibrodzka2014observational},\cite{godel1949example}).

The work is organized as follows. In sections II and III we give a summary of \ac{kds} and \ac{rkds} space-times. We provide analytic solutions to their horizons and radii of equatorial circular photon orbits. We further analyze the behaviour of the radii of the equatorial circular photon orbits in each case. In section IV, we obtain the celestial coordinates for the shadow cast by a \ac{kds} and \ac{rkds} black hole. We analyze the shadows for different values of black hole spin - the parameter $a$ - and cosmological constant $\Lambda$. In section V, we numerically evaluate the curvature radii of the shadows at specific characteristic points. We further obtain numerical values for the horizontal and vertical diameters of the shadows. In section VI, we use the radius of curvature to constrain a \ac{kds} and \ac{rkds} black hole. Lastly, in section VII, we give a conclusion for our results. The metric signature in this work is $- + + +$ and $G=c=1$.

\section{Kerr-de Sitter space time}
\ac{kds} metric is a limited case of the Plebanski-Demianski solution with zero acceleration, NUT parameter, electric and magnetic charges. In Boyer Lindquist coordinates, \ac{kds} space time is characterized by a metric of the form \cite{novikov1973houches},
\begin{multline}
ds^{2}= \left(\frac{a^2 \Delta_{\theta } \sin ^2(\theta )}{L^2 \Sigma }-\frac{\text{$\Delta_{r} $}}{L^2 \Sigma }\right) \text{dt}^2 -\frac{2 a \sin ^2(\theta ) \left(\Delta_{\theta }  \left(a^2+r^2\right)-\text{$\Delta_{r}$}\right)}{L^2 \Sigma } \text{dt} \text{d$\phi $} \\+\frac{ \sin ^2(\theta ) \left(\Delta_{\theta }  \left(a^2+r^2\right)^2-a^2 \text{$\Delta_{r} $} \sin ^2(\theta )\right)}{L^2 \Sigma } \text{d$\phi $}^2+\frac{ \Sigma }{\Delta_{\theta }  } \text{d$\theta $}^2+\frac{ \Sigma }{\text{$\Delta_{r} $}} \text{dr}^2, \label{1}
\end{multline}
where the terms appearing in the metric coefficients are defined as,
\begin{eqnarray}
    \Delta_{\theta}&=&1+\dfrac{\Lambda a^2 \cos^2 \theta}{3}, \\
    \Delta_{r}&=& (1-\dfrac{\Lambda r^{2}}{3})(r^{2}+a^{2})-2M r, \\
    L&=& 1+\dfrac{\Lambda a^2}{3}, \\
    \Sigma &=& r^{2}+a^{2}\cos^{2}\theta.
\end{eqnarray}
Coordinates $t$ and $r$ range over all $\mathbb{R}$ while $\theta \in [0,\pi]$ and $\phi \in [0,2\pi]$. $M$ is the total mass of the system, $a$ the angular momentum per unit mass and $\Lambda$ is the cosmological constant. The coefficients of metric \cref{1} are independent of $t$ and $\phi$ thus $\partial_{t}$ and $\partial_{\phi}$ are Killing vector fields.
Any linear combination of these two Killing vector fields will also be a Killing vector,\cite{visser2007kerr}.
\begin{itemize}
    \item The region where $g_{tt}>0$ is the ergoregion. In the ergoregion, $\partial_{t}$ is space-like. The boundary of the ergoregion occurs at $g_{tt}=0$ which is known as the static limit. The static limit defines how close to the black hole static observers (observers for which spatial coordinates along their world lines do not change with time) can get. In the ergoregion, trajectories for which the observer remains at fixed values of Boyer-Lindquist coordinates are not possible. Observers in this region remain at fixed $r$ and $\theta$ by rotating in $\phi$ direction.
    \item For $g_{\phi \phi}<0$, $\partial_{\phi}$ becomes time-like. This will allow the presence of closed time-like curves whose existence violates causality. Observers moving on closed time-like curves find themselves in their own past. Closed time-like curves were first discovered by Kurt Godel \cite{godel1949example}.
\end{itemize}
Moreover, the metric has coordinate singularities at $\Delta_{r}=0$, $\theta=0$, $\pi$ and a curvature singularity at $\Sigma=0$ (i.e $r=0$ and $\theta=\pi/2$). Kerr-de Sitter horizons are determined by the roots of $\Delta_{r}$, thus,
\begin{align}
(1-\dfrac{\Lambda r^{2}}{3})(r^{2}+a^{2})-2M r=0. \label{delh}
\end{align}
\autoref{delh} is a quartic polynomial hence admits four roots $r_{--}$, $r_{h-}$, $r_{h+}$ and $r_{c}$, with $0$, $2$ or $4$ of the roots being real. $r_{h-}$ and $r_{h+}$ are the Cauchy and Event horizons respectively while $r_{c}$ is the cosmological horizon and $r_{--}$
is interpreted as the dual of the cosmological horizon. Applying Ferrari's solution for quartic polynomials, we obtain the relations for the horizons as,
\begin{align}
r_{--} = -z -\sqrt{\frac{3 \left(1-y\right)}{2 \Lambda }+\frac{3 M}{2 \Lambda  z}-z^2}, \label{horizon1kd} \\
r_{h-} = -z+\sqrt{\frac{3 \left(1-y\right)}{2 \Lambda }+\frac{3 M}{2 \Lambda  z}-z^2}, \label{horizon2kd} \\
r_{h+} = z-\sqrt{\frac{3 \left(1-y\right)}{2 \Lambda }-\frac{3 M}{2 \Lambda  z}-z^2}, \label{horizon3kd} \\
r_{c} = z+\sqrt{\frac{3 \left(1-y\right)}{2 \Lambda }-\frac{3 M}{2 \Lambda  z}-z^2}, \label{horizon4kd}
\end{align}
 where we have defined,
\begin{align}
z = \sqrt{\frac{1}{2} \left(-\frac{a^2}{3}+\frac{1}{\Lambda }+\text{$x_{-} $}+\text{$x_{+} $}\right)}, y=\frac{a^{2}\Lambda}{3},
\end{align}
\begin{align}
x_{\pm} =\frac{y (y (y+33)-33)-1+18 \Lambda  M^2\pm6 \Lambda ^3 \sqrt{\frac{\omega}{\Lambda ^6}}}{8 \Lambda ^3},\\
\omega=9 \Lambda ^2 M^4+\Lambda  M^2 (y-1) (y (y+34)+1)+3 y (y+1)^4. \label{D}
\end{align}
The condition that horizons (\ref{horizon1kd}) - (\ref{horizon4kd}) have to obey to be real is $\omega>0$.
Thus, $\omega>0$ is the condition for a regular \ac{kds} black hole.

\subsection{Photon Region}
Due to the symmetries of \ac{kds} space-time, the trajectories possesses the conserved quantities,
\begin{align}
    p_{t}=-g_{tt} p^{t}-g_{t\phi}p^{\phi} =: E, p_{\phi}=g_{t\phi}p^{t}+g_{\phi \phi}p^{\phi}=:\Phi.
\end{align}
$E$ is interpreted as the energy of the particles per unit mass and is related to the space-time being stationary while $\Phi$ is the angular momentum per unit mass in the $z$ direction and is related to axial symmetry. There exist a third conserved quantity related to the hidden symmetry of \ac{kds} space-time which is the Carter's constant $Q$ \cite{hackmann2010analytical}. $Q$ comes up as a result of separation of variables in the Hamilton-Jacobi equation. For photon orbits, only the sign of energy has a physical meaning. Hence the conserved quantities can be rescaled in terms of energy as,
\begin{align}
    \eta = \frac{Q}{E^{2}}, \lambda=\frac{\Phi}{E}.
\end{align}
Thus the equations describing motion of photon orbits in \ac{kds} space-times are given by the Carter's equations \cite{novikov1973houches},
	\begin{align}
\dfrac{\Sigma}{E}p^{r}=\pm\sqrt{R(r)}, \label{33}\\
\dfrac{\Sigma}{E}p^{\theta}=\pm \sqrt{\Theta(\theta)}, \label{34}\\
\dfrac{\Sigma}{E}p^{\phi}=\dfrac{aL^{2}}{\Delta_{r}}(a(a-\lambda)+r^{2})-\dfrac{L^{2}}{\Delta_{\theta}\sin^{2}\theta}(a\sin^{2}\theta-\lambda) \label{35},\\
\dfrac{\Sigma}{E}p^{t}=\dfrac{L^{2}}{\Delta_{r}}((r^{2}+a^{2})^{2}-a\lambda(a^{2}+r^{2}))-\dfrac{aL^{2}}{\Delta_{\theta}}(a\sin^{2}\theta-\lambda). \label{36}
\end{align}
In the above expressions we have defined,
\begin{align}
    R(r)=L^{2}(r^{2}+a^{2}-a\lambda)^{2}-\Delta_{r}(\eta+L^{2}(\lambda-a)^{2}), \label{37}
\end{align}
\begin{align}
\Theta(\theta)=a^2 \Delta_{ \theta}  L^2+a^2 L^2 \cos ^2(\theta )-a^2 L^2-2 a \Delta_{ \theta } \lambda  L^2+2 a \lambda  L^2+\Delta_{ \theta}  \eta +\Delta_ {\theta}  \lambda ^2 L^2-\lambda ^2 L^2 \cot ^2(\theta )-\lambda ^2 L^2, \label{38} 
\end{align}
with usual definitions $p^{\mu} =dx^{\mu}/d\sigma$, where $\sigma$ is the affine parameter along the geodesic.

An important class of photon orbits pivotal to the formation of a black hole shadow are the spherical photon orbits. 
For such orbits, the conditions,
\begin{align}
R(r)=0, \quad R'(r)=0, \label{circular condition kd}
\end{align}
must be satisfied. Solving both conditions in \cref{circular condition kd} simultaneously yields,
\begin{align}
\eta = -\frac{L^2 r^3 \left(6 a^2 \left(\Lambda  r^2 (3 M+r)-6 M\right)+a^4 \Lambda ^2 r^3+9 r (r-3 M)^2\right)}{a^2 \left(r \left(a^2 \Lambda +2 \Lambda  r^2-3\right)+3 M\right)^2}, \label{eta kd} \\
\lambda = \frac{r \left(a^2 \left(6-\Lambda  r^2\right)+3 r (r-3 M)\right)}{a \left(r \left(a^2 \Lambda +2 \Lambda  r^2-3\right)+3 M\right)}+a. \label{lambda kd}
\end{align}
In the case $a=0$, \cref{circular condition kd} reduces to \ac{sds}. Due to spherical symmetry of \ac{sds} space-time, orbits will be planar hence we can choose the plane $\theta=\pi/2$. On this plane, $\eta=0$. Thus substituting for ($a=0$, $\eta=0$) in \cref{circular condition kd}, we obtain the relation for $\lambda$ in \ac{sds} space-time as \cite{perlick2022calculating},
\begin{align}
    \lambda_{SdS}=\frac{\sqrt{6} r^{3/2}}{\sqrt{-3 M-2 \Lambda  r^3+3 r}}. \label{lambda sds}
\end{align}

Setting $\Lambda=0$ in \cref{eta kd} and (\ref{lambda kd}) yields parameters obtained in Kerr space-time \cite{cunha2018shadows}, 
\begin{align}
    \Bar{\eta}=-\frac{r^3 \left(r (r-3 M)^2-4 a^2 M\right)}{a^2 (M-r)^2}, \\
    \Bar{\lambda}= \frac{a^2 (M+r)+r^2 (r-3 M)}{a (M-r)}.
\end{align}
\autoref{eta kd} and (\ref{lambda kd}) are the relations for the constants of motion governing photon orbits at $r=$ constant. The radii of these orbits are bound by the equatorial circular photon orbits which are obtained by the condition $\eta=0$. We note that solving for $\eta=0$ is a cubic polynomial hence results in three roots. However, we will give the two roots that are relevant for this work (roots located outside the event horizon). By using Cardano's formula, the roots read,

\begin{align}
r_{ph+,KdS}=-\frac{2 M (y-1)}{(y+1)^2}+2 \sqrt{\frac{M^2 ((y-14) y+1)}{(y+1)^4}}  \cos \left(\frac{\kappa }{3}+\frac{4 \pi }{3}\right), \label{3}\\
r_{ph-,KdS}=-\frac{2 M (y-1)}{(y+1)^2}+2 \sqrt{\frac{M^2 ((y-14) y+1)}{(y+1)^4}}  \cos \left(\frac{\kappa }{3}\right). \label{4}
\end{align}
 The parameter $\kappa$ is defined via the following expression,
\begin{align}
\kappa=\arccos\left( \frac{M \left(2 a^2 (y+1)^4+M^2 (y-1) (y (y+34)+1)\right)}{(y+1)^6 \left(\frac{M^2 ((y-14) y+1)}{(y+1)^4}\right)^{3/2}} \right). \label{kappa}
\end{align}
Thus the photon region in \ac{kds} black hole exterior exist in the range $r\in [r_{ph+,KdS},r_{ph-,KdS}]$. Photon orbits at $r=$constant are unstable with respect to radial perturbations if $d^{2}R/dr^{2}>0$ and stable if $d^{2}R/dr^{2}<0$.

\begin{figure}[htb]
	\centering 
	\begin{subfigure}{0.5\textwidth}
		\includegraphics[width=1.0\linewidth]{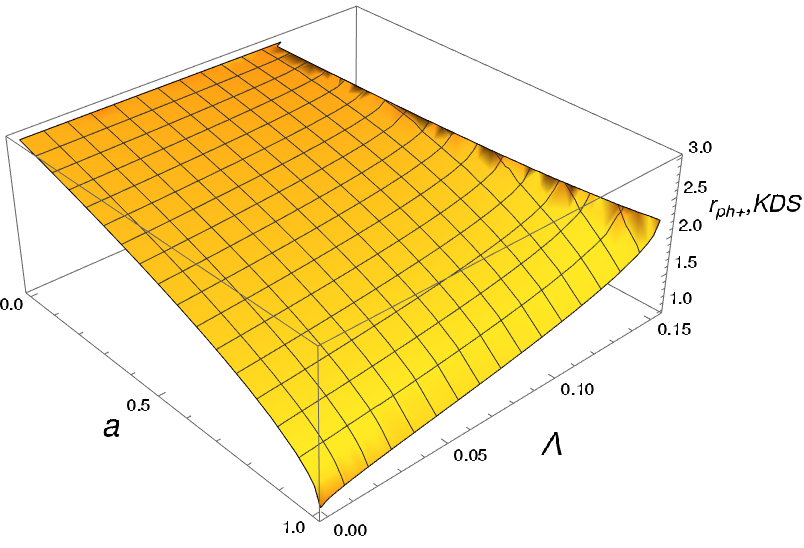}
		\caption{$r_{ph+,KdS}$.}
		\label{fig:rppkd}
	\end{subfigure}\hfil
	\begin{subfigure}{0.5\textwidth}
		\includegraphics[width=1.0\linewidth]{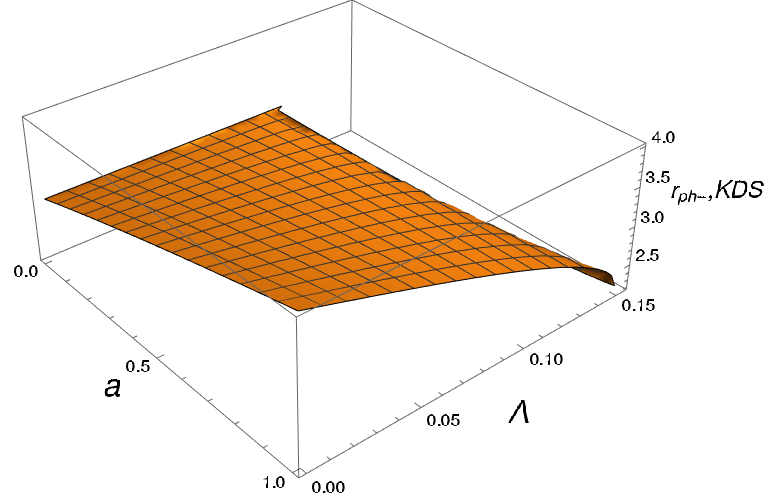}
		\caption{$r_{ph-,KdS}$.}
		\label{fig:rpmkd}
	\end{subfigure}\hfil
	
	\caption{Behaviour of radius of the equatorial circular photon orbit, $r_{ph-,KdS}$ and $r_{ph+,KdS}$ , for $a\in [0,1]$ and $\Lambda\in [0,\Lambda_{Max,KdS}]$. As the black hole spin increases, $r_{ph+,KdS}$ decreases while $r_{ph-,KdS}$ increases. On the other hand, as $\Lambda$ increases, $r_{ph+,KDS}$ increases while $r_{ph-,KdS}$ decreases. We expect to observe this behaviour on the \ac{kds} black hole shadows for varying values of $\Lambda$ or $a$.
	}
	\label{fig:rpkd}
\end{figure}

\autoref{fig:rpkd} illustrates the behaviour of the radius of equatorial circular photon orbits, \cref{3} and (\ref{4}).

\section{Kerr-de Sitter Revisited space-time}
The \ac{rkds} solution has been recently proposed by  Ovalle, Contreras \& Stuchlik in Ref. \cite{ovalle2021kerr} and is defined by the metric, 
\begin{align}
ds^{2}=-\left( \dfrac{\Delta_{\Lambda}-a^{2}\sin^{2}\theta}{\rho^{2}}\right) dt^{2}+\dfrac{\rho^{2}}{\Delta_{\Lambda}} dr^{2}+\rho^{2}d\theta^{2}+\dfrac{\Sigma_{\Lambda}\sin^{2}\theta}{\rho^{2}}d\phi^{2}-\dfrac{2a\sin^{2}\theta}{\rho^{2}}(r^{2}+a^{2}-\Delta_{\Lambda})dt d\phi, \label{kdsrevisited}
\end{align}
with the following definitions,
\begin{align}
\Delta_{\Lambda}=r^{2}-2 M r+a^{2}-\dfrac{\Lambda r^{4}}{3}, \\
\Sigma_{\Lambda}=(r^{2}+a^{2})^{2}-\Delta_{\Lambda}a^{2}\sin^{2}\theta,\\
\rho^{2}=r^{2}+a^{2} \cos^{2}\theta.
\end{align}
In this solution, $\Lambda$ is interpreted as a vacuum energy that suffers distortions in the vicinity of the black hole due to rotation.
Coordinates $r$, $t$ $\in$ $\mathbb{R}$ while $\theta \in [0,\pi]$ and $\phi \in [0,2\pi)$. The metric is stationary and axially symmetric (i.e $\partial_{t}$ and $\partial_{\phi}$ are Killing vector fields). The geometric interpretation for metric components $g_{tt}$ and $g_{\phi\phi}$ revised in section II can be immediately applied now. From the condition,
\begin{align}
r^{2}-2 M r+a^{2}-\dfrac{\Lambda r^{4}}{3} =0, \label{rkdshorizons}
\end{align}
making use of Ferrari's solution we obtain the horizons as of the \ac{rkds} metric such that,
\begin{align}
    \tilde{r}_{--} = -\frac{\sqrt{\frac{1}{\Lambda }+\text{$\varrho_{-} $}+\text{$\varrho_{+} $}}}{\sqrt{2}}-\sqrt{\frac{1}{\Lambda }+\frac{1}{2} \left(\frac{3 \sqrt{2} M}{\Lambda  \sqrt{\frac{1}{\Lambda }+\text{$\varrho_{-} $}+\text{$\varrho_{+} $}}}-\text{$\varrho_{-} $}-\text{$\varrho_{+} $}\right)}, \label{horizon1kdr} \\
    \tilde{r}_{h-} =-\frac{\sqrt{\frac{1}{\Lambda }+\text{$\varrho_{-}$}+\text{$\varrho_{+} $}}}{\sqrt{2}}+ \sqrt{\frac{1}{\Lambda }+\frac{1}{2} \left(\frac{3 \sqrt{2} M}{\Lambda  \sqrt{\frac{1}{\Lambda }+\text{$\varrho_{-} $}+\text{$\varrho_{+} $}}}-\text{$\varrho_{-} $}-\text{$\varrho_{+} $}\right)}, \label{horizon2kdr} \\
    \tilde{r}_{h+} = \frac{\sqrt{\frac{1}{\Lambda }+\text{$\varrho_{-} $}+\text{$\varrho_{+} $}}}{\sqrt{2}} -\sqrt{\frac{1}{\Lambda }+\frac{1}{2} \left(-\frac{3 \sqrt{2} M}{\Lambda  \sqrt{\frac{1}{\Lambda }+\text{$\varrho_{-} $}+\text{$\varrho_{+} $}}}-\text{$\varrho_{-} $}-\text{$\varrho_{+} $}\right)}, \label{horizon3kdr} \\
    \tilde{r}_{c} = \frac{\sqrt{\frac{1}{\Lambda }+\text{$\varrho_{-} $}+\text{$\varrho_{+} $}}}{\sqrt{2}}+\sqrt{\frac{1}{\Lambda }+\frac{1}{2} \left(-\frac{3 \sqrt{2} M}{\Lambda  \sqrt{\frac{1}{\Lambda }+\text{$\varrho_{-} $}+\text{$\varrho_{+} $}}}-\text{$\varrho_{-} $}-\text{$\varrho_{+} $}\right)}. \label{horizon4kdr}
\end{align}

The above results made use of the definitions,
\begin{align}
   \varrho_{\pm} = \left[\frac{1}{2} \left(\frac{9 M^2}{2 \Lambda ^2}-\frac{\frac{3 a^2}{\Lambda }+\frac{1}{4 \Lambda ^2}}{\Lambda }\right) \pm\sqrt{\frac{1}{27} \left(\frac{3 a^2}{\Lambda }-\frac{3}{4 \Lambda ^2}\right)^3+\frac{1}{4} \left(\frac{\frac{3 a^2}{\Lambda }+\frac{1}{4 \Lambda ^2}}{\Lambda }-\frac{9 M^2}{2 \Lambda ^2}\right)^2} \right]^{1/3}.
\end{align}
Whereas $\tilde{r}_{h+}$ and $r_{h-}$ are the event horizon and Cauchy horizon respectively, $\tilde{r}_{c}$ is the cosmological horizon and $\tilde{r}_{--}$ is interpreted as the dual of the cosmological horizon.
These roots will be real provided the discriminant of \cref{rkdshorizons} is positive,

\begin{align}
-\frac{27}{4 \Lambda ^5}(16 a^6 \Lambda ^2+24 a^4 \Lambda +a^2 \left(9-108 \Lambda  M^2\right)+81 \Lambda  M^4-9 M^2) >0. \label{discriminantkdr}
\end{align}
Thus \cref{discriminantkdr} is the condition for the existence of a regular \ac{rkds} black hole.
\subsection{Photon Region}
\begin{figure}[htb]
	\centering 
	\begin{subfigure}{0.5\textwidth}
		\includegraphics[width=1.0\linewidth]{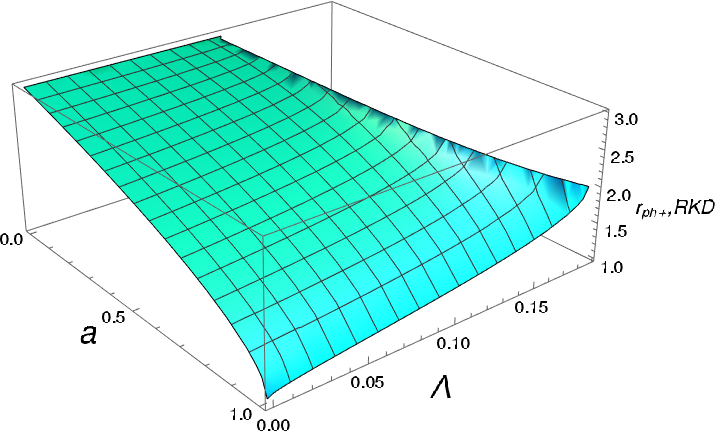}
		\caption{$r_{ph+,RKdS}$.}
		\label{fig:rpprkd}
	\end{subfigure}\hfil
	\begin{subfigure}{0.5\textwidth}
		\includegraphics[width=1.0\linewidth]{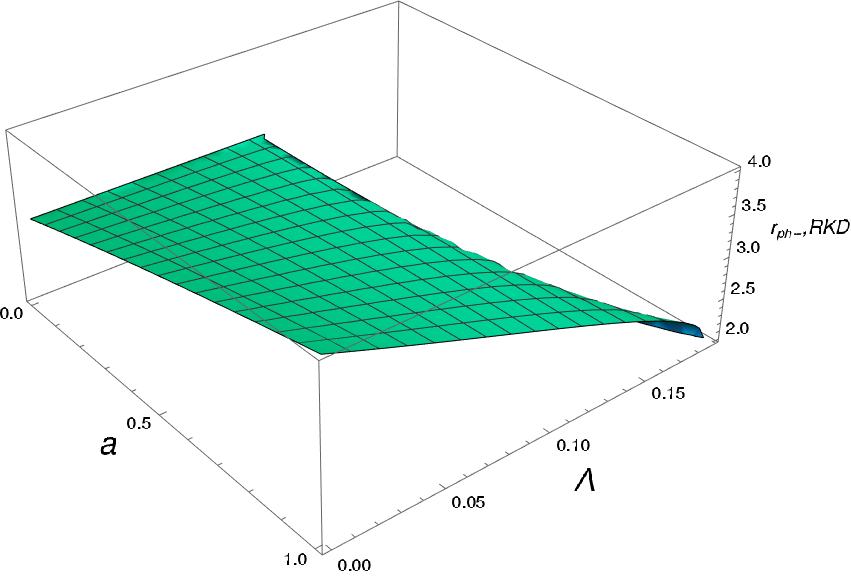}
		\caption{$r_{ph-,RKdS}$.}
		\label{fig:rpmrkd}
	\end{subfigure}\hfil
	
	\caption{Behaviour of radius of the equatorial circular photon orbit, $r_{ph-,RKdS}$ and $r_{ph+,RKdS}$ , for $a\in [0,1]$ and $\Lambda\in [0,\Lambda_{Max,RKdS}]$. As the black hole spin increases, $r_{ph+,RKdS}$ decreases while $r_{ph-,RKdS}$ increases. On the other hand, as $\Lambda$ increases, $r_{ph+,RKdS}$ increases while $r_{ph-,RKdS}$ decreases. We expect to observe this behaviour on the shadows of \ac{rkds} black hole for varying $\Lambda$ or $a$.
	}
	\label{fig:rprkd}
\end{figure}

Due to \ac{rkds} solution being stationary and axially symmetric, we have the conserved quantities,
\begin{align}
    p_{t,\Lambda}=-g_{tt,\Lambda} p^{t}_{\Lambda}-g_{t\phi,\Lambda}p^{\phi}_{\Lambda} =: E_{\Lambda}, p_{\phi,\Lambda}=g_{t\phi,\Lambda}p^{t}_{\Lambda}+g_{\phi \phi,\Lambda}p^{\phi}_{\Lambda}=:\Phi_{\Lambda}.
\end{align}
Where $E_{\Lambda}$ is the energy per unit mass and $\Phi_{\Lambda}$ is the angular momentum in the $z$ direction. Despite \ac{rkds} solution being a non-vacuum solution, we find that the Hamilton-Jacobi equation is separable and thus this space-time possesses the Carter's constant $Q_{\Lambda}$, 
\begin{align}
    Q_{\Lambda}= p_{\theta, \Lambda}^{2}-a^{2} p_{t,\Lambda}^{2} \cos^{2}\theta+p_{\phi, \Lambda}^{2}\cot^2 \theta. \label{carter rkds}
\end{align}
For photons, we can rescale the conserved quantities in terms of energy as,
\begin{align}
    \lambda_{\Lambda}=\Phi_{\Lambda}/E_{\Lambda},
    \eta_{\Lambda} = Q_{\Lambda}/E_{\Lambda}^{2}.
\end{align}
Using the Hamilton-Jacobi formalism and making use of these symmetries, we obtain the equations for null geodesics as,
\begin{align}
\dfrac{\rho^{2}}{E} p^{r}_{\Lambda}=\pm \sqrt{R_{\Lambda}(r)}, \label{82} \\
\dfrac{\rho^{2}}{E} p^{\theta}_{\Lambda}=\pm \sqrt{\Theta_{\Lambda}(\theta)} \label{83},\\
\dfrac{\rho^{2}}{E}p^{\phi}_{\Lambda}=\dfrac{(a r^{2}+a^{3}-a\Delta_{\Lambda}-a^{2}\lambda_{\Lambda})}{\Delta_{\Lambda}}+\dfrac{\lambda_{\Lambda}}{\sin^{2}\theta}, \label{84}\\
\dfrac{\rho^{2}}{E}p^{t}_{\Lambda}= \dfrac{(r^{2}+a^{2})(r^{2}+a^{2}-a \lambda_{\Lambda})}{\Delta_{\Lambda}}+ a\lambda_{\Lambda}-a^{2}\sin^{2}\theta, \label{85}
\end{align}
with the following definitions,
\begin{align}
R_{\Lambda}(r)=(r^{2}+a^{2}-a\lambda_{\Lambda})^{2}-\Delta_{\Lambda}(\eta_{\Lambda}+(\lambda_{\Lambda}-a)^{2}), \\
\Theta_{\Lambda}(\theta)=\eta_{\Lambda}+a^{2}\cos^{2}\theta-\lambda^{2}_{\Lambda} \cot^{2}\theta.\end{align}
Also, it is worth noting that
 we proceed as before and by simultaneously solving $R_{\Lambda}(r)=R'_{\Lambda}(r)=0$, which results in the following relations,
\begin{align}
\eta_{\Lambda} =-\frac{3 r^3 \left(4 a^2 \left(\Lambda  r^3-3 M\right)+3 r (r-3 M)^2\right)}{a^2 \left(3 M+2 \Lambda  r^3-3 r\right)^2}, \label{88a}\\
\lambda_{\Lambda} =\frac{3 a^2 M+2 a^2 \Lambda  r^3+3 a^2 r-9 M r^2+3 r^3}{a \left(3 M+2 \Lambda  r^3-3 r\right)}. \label{89}
\end{align}
The \ac{rkds} space-time reduces to \ac{sds} in the limit $a=0$. Substituting for ($a=0$ $\eta_{\Lambda}=0$) in $R_{\Lambda}(r)=R'_{\Lambda}(r)=0$ and solving for $\lambda$ then gives \cite{perlick2022calculating}, 
\begin{align}
    \lambda_{SdS, \Lambda}=\frac{\sqrt{6} r^{3/2}}{\sqrt{-3 M-2 \Lambda  r^3+3 r}}. \label{rsds}
\end{align}
$\eta_{\Lambda}=0$ since we chose the plane $\theta=\pi/2$ as a result of \ac{sds} space-time being spherically symmetric.
For $\Lambda=0$, \cref{88a} and (\ref{89}) yields equations same as those obtained in Kerr space-time \cite{cunha2018shadows}, 
\begin{align}
    \Bar{\eta_{\Lambda}}=-\frac{r^3 \left(r (r-3 M)^2-4 a^2 M\right)}{a^2 (M-r)^2}, \\
    \Bar{\lambda_{\Lambda}}= \frac{a^2 (M+r)+r^2 (r-3 M)}{a (M-r)}.
\end{align}
\autoref{88a} and (\ref{89}) are the relations for constants  governing motion of photon orbits at constant $r$ in \ac{rkds} space-time. 
The zeros of \cref{88a}, yields the radii of equatorial circular photon orbits,
\begin{align}
r_{ph+,RKdS} = \frac{6 M}{4 a^2 \Lambda +3} +6 \sqrt{\frac{M^2 \left(1-4 a^2 \Lambda \right)}{\left(4 a^2 \Lambda +3\right)^2}} \cos \left(\frac{\tilde{\kappa} }{3}+\frac{4 \pi }{3}\right), \label{rphp kdr} \\
r_{ph-,RKdS} = \frac{6 M}{4 a^2 \Lambda +3} +6 \sqrt{\frac{M^2 \left(1-4 a^2 \Lambda \right)}{\left(4 a^2 \Lambda +3\right)^2}} \cos \left(\frac{\tilde{\kappa} }{3}\right). \label{rphm kdr}
\end{align}
$\tilde{\kappa}$ is defined via the expression,
\begin{align}
\tilde{\kappa} = \arccos\left( \frac{9 M^2-2 a^2 \left(2 \Lambda  \left(8 a^4 \Lambda +12 a^2-27 M^2\right)+9\right)}{9 M \left(4 a^2 \Lambda -1\right) \left(4 a^2 \Lambda +3\right) \sqrt{\frac{M^2 \left(1-4 a^2 \Lambda \right)}{\left(4 a^2 \Lambda +3\right)^2}}}\right). 
\end{align}
Thus the radii of photon orbits with $r=$constant is bound by $r \in [r_{ph+,RKdS},r_{ph-,RKdS}]$. These orbits are unstable to radial perturbations when $R''_{\Lambda}>0$ and stable if $R''_{\Lambda}<0$. \autoref{fig:rprkd} illustrates the behaviour of the radius of equatorial circular photon orbits, \cref{rphp kdr}  and (\ref{rphm kdr}). 

 Up to this point we have extensively compared the \ac{kds} and \ac{rkds} space times from the qualitative point of view. Henceforth we investigate quantitatively  the predictions for the black hole shadows in both space-times and compare them.

\section{Kerr-de Sitter and Kerr-de Sitter Revisited black hole shadows}
Utilizing the null geodesics of \ac{kds} space-time, we obtain the corresponding impact parameters as,

\begin{align}
\alpha_{KdS}=\frac{\sqrt{3} L^2 \sin (\theta ) \left(a-\lambda  \csc ^2(\theta )\right)}{\Delta_{\theta}  \sqrt{L^2 \left(a^2 \Lambda -2 a \lambda  \Lambda +\lambda ^2 \Lambda +3\right)+\eta  \Lambda }}, \label{50} \\
\beta_{KdS}= \frac{\sqrt{3}(\pm \sqrt{\Theta (\theta )})}{\sqrt{L^2 \left(a^2 \Lambda -2 a \lambda  \Lambda +\lambda ^2 \Lambda +3\right)+\eta  \Lambda }}, \label{51}
\end{align}
while from the null geodesics of \ac{rkds} space-time we obtain,
\begin{align}
\alpha_{RKdS} =-\frac{\sqrt{3} \csc (\theta ) (a \cos (2 \theta )-a+2 \lambda_{\Lambda} )}{2 \sqrt{a^2 \Lambda -2 a \lambda_{\Lambda}  \Lambda +\eta_{\Lambda}  \Lambda +\lambda^{2}_{\Lambda}  \Lambda +3}}, \label{93} \\
\beta_{RKdS} =\frac{\pm\sqrt{3} \sqrt{a^2 \cos ^2(\theta )+\eta_{\Lambda} -\lambda ^{2}_{\Lambda} \cot ^2(\theta )}}{\sqrt{a^2 \Lambda -2 a \lambda_{\Lambda}  \Lambda +\eta_{\Lambda}  \Lambda +\lambda ^{2}_{\Lambda} \Lambda +3}}. \label{94}
\end{align}

\begin{figure}[htb]
	\centering 
	\begin{subfigure}{0.5\textwidth}
		\includegraphics[width=1.0\linewidth]{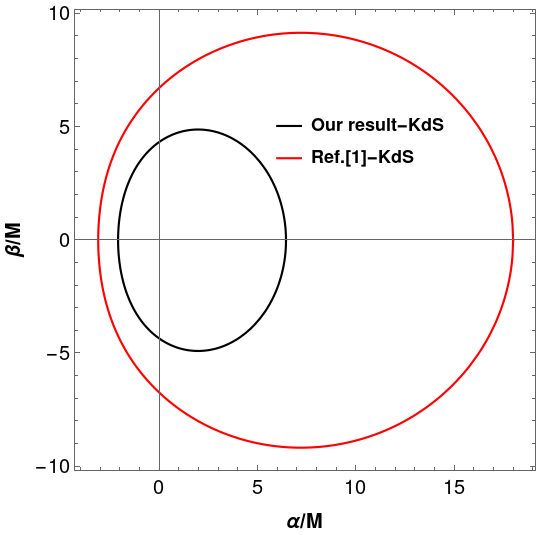}
		\caption{Comparison of the impact parameters in \cref{50} and (\ref{51}) with those in \cref{204} and (\ref{205}).}
		\label{fig:kdsjust}
	\end{subfigure}\hfil
	\begin{subfigure}{0.485\textwidth}
		\includegraphics[width=1.0\linewidth]{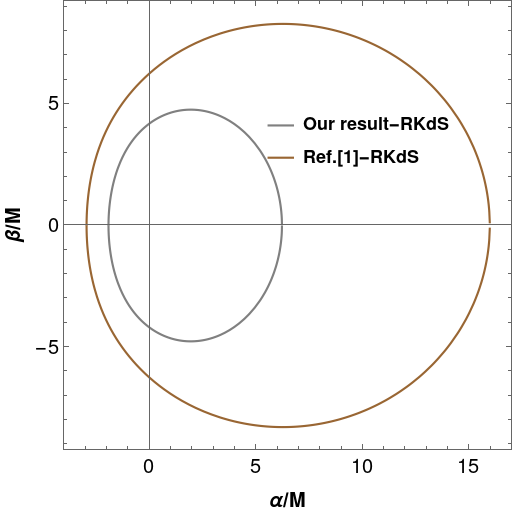}
		\caption{Comparison of the impact parameters in \cref{93} and (\ref{94}) with those in \cref{204} and (\ref{205}).}
		\label{fig:rkdsjust}
	\end{subfigure}\hfil
	
	\caption{Illustration of the discrepancy between our result and Ref. \cite{ovalle2021kerr}. $\Lambda=0.06m^{-2}$, $M=1.06$,$\theta=\pi/2$, and  $a=0.999$.	}
	\label{fig:bothjust}
\end{figure}

We however note that our results \cref{50} and (\ref{51}), \cref{93} and (\ref{94}), are different from those used for  in Ref.\cite{ovalle2021kerr}, where in our observation, for both \ac{kds} and \ac{rkds}, they consider impact parameters of the form \cite{vazquez2003strong},
\begin{align}
\alpha=-\lambda \csc \theta, \label{204}\\
\beta=\pm \sqrt{\eta+a^{2}\cos^{2}\theta-\lambda^{2}\cot^{2}\theta}\label{205}.
\end{align}
Thus, for the apparent displacement of the image perpendicular to the projected axis of symmetry of the black hole, we obtain the impact parameter as in \cref{50} and \cref{93} for \ac{kds} and \ac{rkds} respectively. These equations are clearly different from \cref{204}. Moreover, for the apparent displacement of the image parallel to the projected axis of symmetry of the black hole, we obtain the impact parameter as in \cref{51} and \cref{94} for \ac{kds} and \ac{rkds} respectively. These equations are also different from \cref{205}. This difference can also be observed as illustrated in \cref{fig:bothjust}. In \cref{fig:bothjust}, for the same values of black hole parameters, our results do not coincide with Ref. \cite{ovalle2021kerr} in both \ac{kds} and \ac{rkds} black holes.\\
The error in the impact parameters considered in Ref. \cite{ovalle2021kerr} is due to the fact that they directly considered the impact parameters of Kerr space-time for which they cited Ref.\cite{vazquez2003strong}. Physically, one should utilize the null geodesic equations of the space-time under consideration in order to obtain the impact parameters. Thus, the impact parameters in Ref.\cite{ovalle2021kerr} have been obtained using null geodesic equations of Kerr space-time and this leads to inconsistent  black hole shadow results when applied to \ac{kds} and \ac{rkds} space-times. We however note that this error is not relevant to Ref.\cite{ovalle2021kerr} as they show a shadow to complement their work, without this being the relevant part of their paper. For the purposes of our work and any other future studies this error is relevant. Hence, our impact parameters yield consistent black hole shadow results, as will be seen in the next sections, because we derived them using the null geodesic equations of \ac{kds} and \ac{rkds} space-time respectively. \\
 The boundary of the shadow is formed by spherical photon orbits that are unstable to radial perturbations. For a \ac{kds} black hole  we found that such orbits lie in the range $r_{ph+,KdS} \leq r \leq r_{ph-,KdS}$ while $r_{ph+,RKdS} \leq r \leq r_{ph-,RKdS}$ for a \ac{rkds} black hole. As a result, the impact parameters will be evaluated for values of $r$ in these ranges. 

\autoref{fig:characpoints} demonstrates a general contour of a black hole shadow. The contour is a parametric curve $(\alpha,\beta)$. Every point along this curve is associated to a particular radial point in the photon region. When the observer is inclined on the equatorial plane $r_{D}=r_{ph+,KdS}(r_{ph+,RKdS})$ and $r_{R}=r_{ph-,KdS}(r_{ph-,RKdS})$, which are the radii of equatorial circular photon orbits. 
$\alpha$ increases from a negative value at $r_{D}$ to a positive value at $r_{R}$. $\beta$ on the other hand vanishes at $D$ and $R$. Thus when an observer is inclined on the equatorial plane,  the behaviour of the shadows at point $D$ and $R$ should correspond to the behaviour of the radii of equatorial circular photon orbits  that we obtained in \cref{fig:rpkd} and \cref{fig:rprkd}. In other words, 
\begin{itemize}
    \item For fixed $\Lambda$, $r_{ph+,KdS}(r_{ph+,RKdS})$ both decrease with increase in black hole spin. This implies that as black hole spin increases, point $D$ on the curve of the shadow will decrease by moving to the right. $r_{ph-,KdS}(r_{ph-,RKdS})$ increases with increase in black hole spin, thus point $R$ increases by moving further to the right. As a consequence, when the black hole spin increases, the shadows will appear to shift to the right.
    \item For fixed black hole spin, $r_{ph+,KdS}(r_{ph+,RKdS})$ increases with increase in $\Lambda$. In this case, point $D$ increases by moving further to the left. $r_{ph-,KdS}(r_{ph-,RKdS})$ decreases with increasing black hole spin. Point $R$ consequently decreases by moving to the left. Hence, with increasing $\Lambda$, the shadows undergo a shift to the left.
\end{itemize}
 This behaviour maintains even when the observer is inclined away from the equatorial plane. In the subsequent section, we will analyze the qualitative behaviour of the shadows using this concept so as to identify consistencies or inconsistencies.
\subsection{Plots of Black Hole shadows}
\begin{figure}[htb]
	\centering 
	\begin{subfigure}{0.5\textwidth}
		\includegraphics[width=0.7\linewidth]{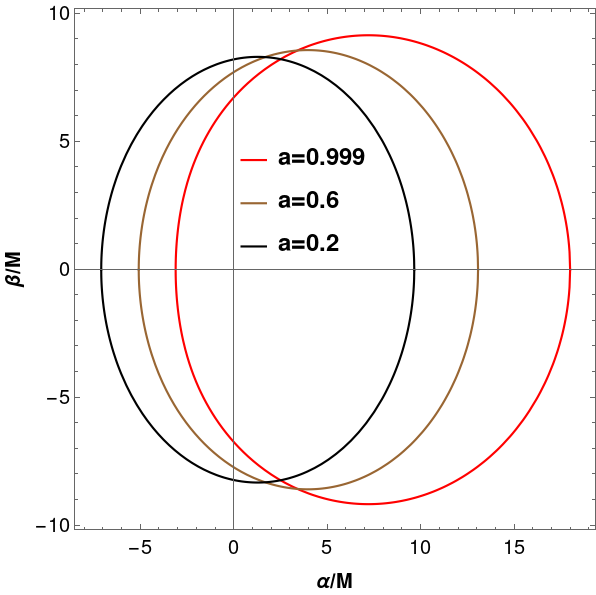}
		\caption{Kerr-De Sitter Black Hole shadow for celestial coordinates in Ref.\cite{ovalle2021kerr}, \cref{204} and (\ref{205}). }
		\label{fig:ovaleda}
	\end{subfigure}\hfil
	\begin{subfigure}{0.5\textwidth}
		\includegraphics[width=0.7\linewidth]{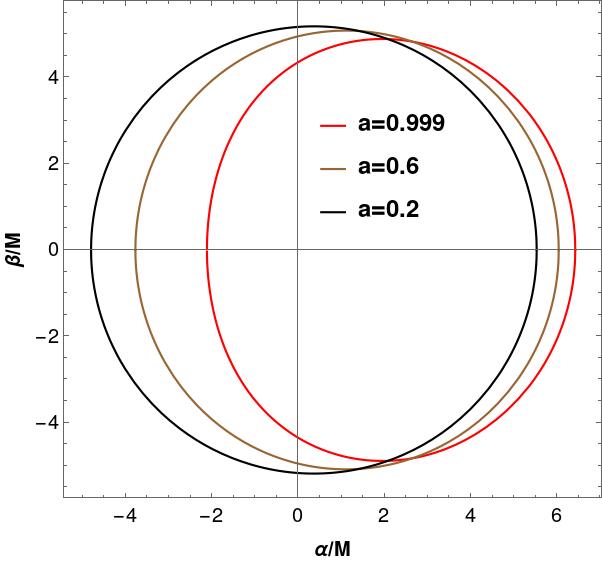}
		\caption{Kerr-De Sitter Black Hole shadow for our celestial coordinates \cref{50} and (\ref{51}).}
		\label{fig:mineovaleda}
	\end{subfigure}\hfil
	\begin{subfigure}{0.5\textwidth}
		\includegraphics[width=0.7\linewidth]{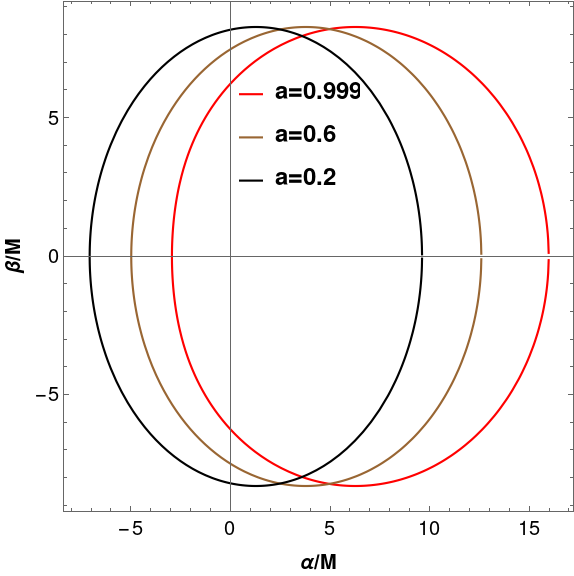}
		\caption{Kerr-De Sitter Revisited Black Hole shadow for our celestial coordinates in Ref.\cite{ovalle2021kerr}, \cref{204} and (\ref{205}). }
		\label{fig:ovaledar}
	\end{subfigure}\hfil
	\begin{subfigure}{0.5\textwidth}
		\includegraphics[width=0.7\linewidth]{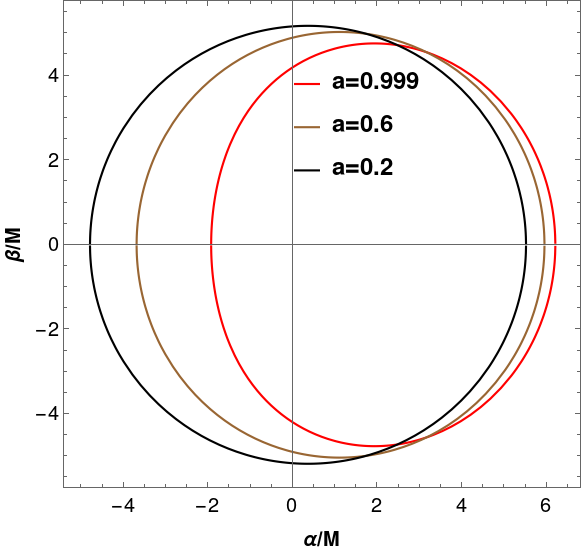}
		\caption{Kerr-De Sitter Revisited Black Hole shadow for celestial coordinates, \cref{93} and (\ref{94}).}
		\label{fig:mineovaledaR}
	\end{subfigure}\hfil
	\caption{Comparison of the shadows for impact parameter used in Ref.\cite{ovalle2021kerr} with those that we obtain. We fix $\Lambda=0.06m^{-2}$, $M=1.06$,$\theta=\pi/2$, and vary $a$.  }
	\label{fig:mineovale}
\end{figure}
In \cref{fig:bothjust}, (\ref{fig:mineovale}) and (\ref{fig:dscomparison}), we have plotted the shadows using a  $\Lambda=0.06$ and $M=1.06$ as considered in Ref. \cite{ovalle2021kerr} (figure 2 of their work). For different values of the black hole spin, $a$, we notice that our results qualitatively agree with Ref.\cite{ovalle2021kerr} i.e, as the black hole spin increases, the left side of the shadows moves to the right which corresponds to the radius of  equatorial circular prograde photon orbit decreasing with increase in black hole spin. The right side moves further to the right which is consistent with the radius of equatorial circular retrograde photon orbit increasing with an increase in black hole spin.
Nonetheless, it is clear that there exists a quantitative discrepancy. The celestial coordinates used in Ref. \cite{ovalle2021kerr} yield shadows that appear elliptic. On the other hand, our results yield shadows that do not deviate so much from circularity. 

\begin{figure}[htb]
	\centering 
	\begin{subfigure}{0.5\textwidth}
		\includegraphics[width=0.7\linewidth]{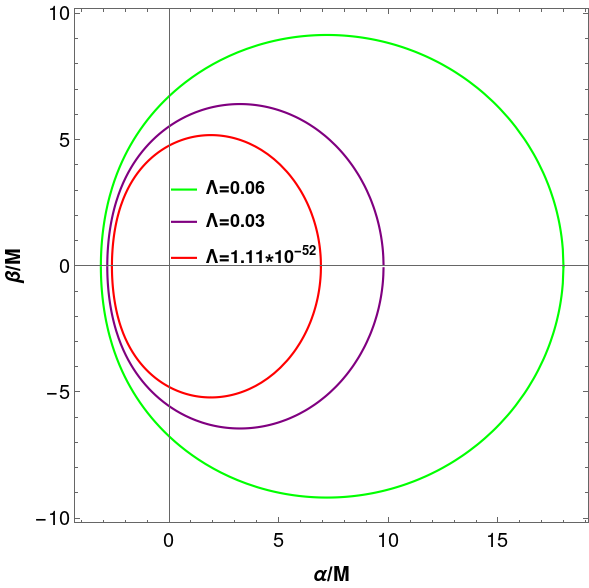}
		\caption{\ac{kds} Black Hole shadow for celestial coordinates in Ref. \cite{ovalle2021kerr}, \cref{204} and (\ref{205}). }
		\label{fig:ovaledt}
	\end{subfigure}\hfil
	\begin{subfigure}{0.5\textwidth}
		\includegraphics[width=0.7\linewidth]{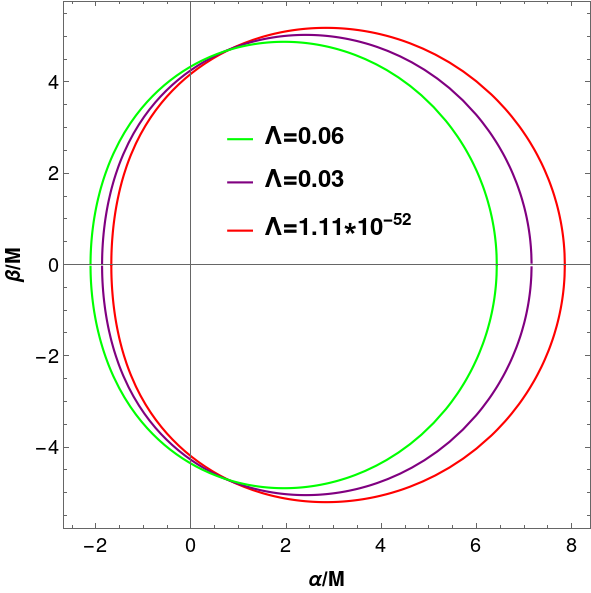}
		\caption{\ac{kds} Black Hole shadow for celestial coordinates \cref{50} and (\ref{51}).}
		\label{fig:mineovaledt}
	\end{subfigure}\hfil
	\begin{subfigure}{0.5\textwidth}
		\includegraphics[width=0.7\linewidth]{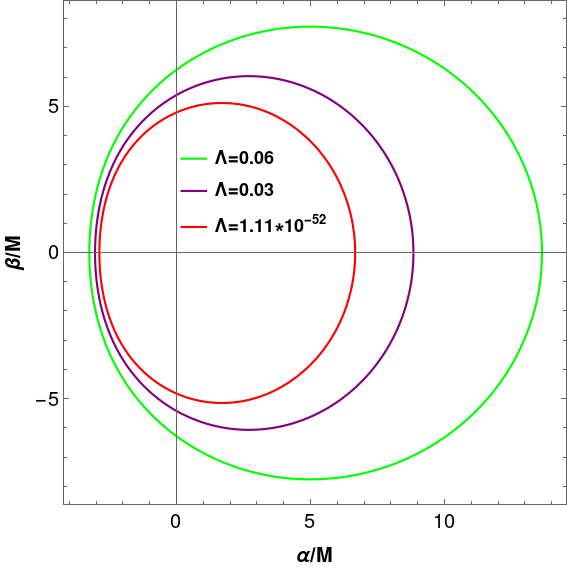}
		\caption{\ac{rkds} Black Hole shadow for celestial coordinates in Ref.\cite{ovalle2021kerr}, \cref{204} and (\ref{205}). }
		\label{fig:ovaledtr}
	\end{subfigure}\hfil
	\begin{subfigure}{0.5\textwidth}
		\includegraphics[width=0.7\linewidth]{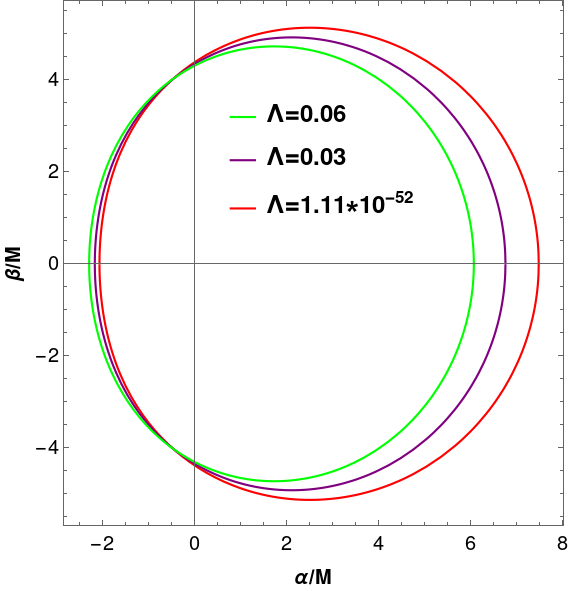}
		\caption{\ac{rkds} Black Hole shadow for celestial coordinates \cref{93} and (\ref{94}).}
		\label{fig:mineovaledtR}
	\end{subfigure}\hfil	
	\caption{Comparison of the shadows for impact parameter used in Ref.\cite{ovalle2021kerr} with those that we obtain. For fixed $a=0.999$, $M=1.06$, $\theta=\pi/2$, $a=0.999$ and different values of $\Lambda$. }
	\label{fig:mineovalet}
\end{figure}
From \cref{fig:mineovalet}, we observe that for increasing $\Lambda$, our obtained impact parameters yield shadows that undergo a shift to the left. This is consistent with the discussion that we have given in the previous section. The impact parameters used in Ref.\cite{ovalle2021kerr}, gives rise to shadows inconsistent with the behaviour of equatorial circular photon orbits, i.e the right side of the shadows moves further to the right as $\Lambda$ increases. This would imply that $r_{ph-,KdS}(r_{ph-,RKdS})$ increases with increase in $\Lambda$ which is not the case as observed in the previous section. In-addition, there is still a quantitative discrepancy.\\
\begin{figure}[htb]
	\centering 
	\begin{subfigure}{0.35\textwidth}
		\includegraphics[width=1.0\linewidth]{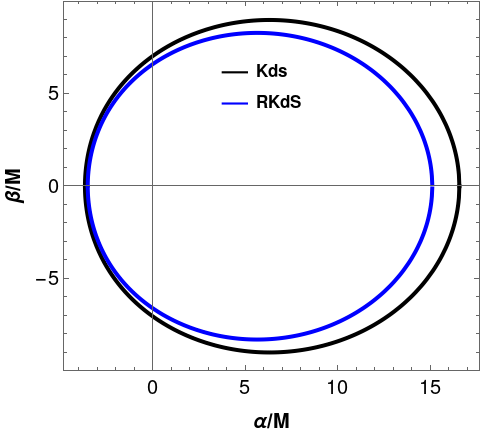}
		\caption{Comparison of \ac{kds} and \ac{rkds} black hole shadow, \cref{204} and (\ref{205}).}
		\label{fig:ORI2}
	\end{subfigure}\hfil
	\begin{subfigure}{0.35\textwidth}
		\includegraphics[width=1.0\linewidth]{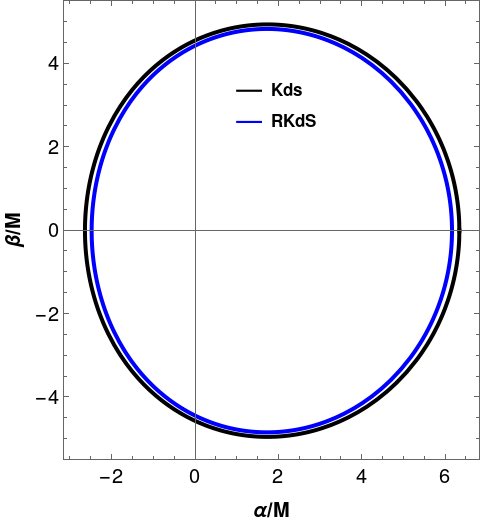}
		\caption{Comparison of a \ac{kds} black hole shadow, \cref{50} and (\ref{51}) with  \ac{rkds} black hole, \cref{93} and (\ref{94})}
		\label{fig:bothkds2}
	\end{subfigure}\hfil
	\caption{Plots comparing the size of a \ac{kds} with \ac{rkds} black hole shadow for $a=0.9$, $M=1.06$, $\theta=\pi/2$ and $\Lambda=0.06m^{-2}$.
	}
	\label{fig:dscomparison}
\end{figure}
\autoref{fig:dscomparison} shows that a \ac{kds} black hole will cast a larger shadow compared to a \ac{rkds} black hole. This observation, as in Ref.\cite{ovalle2021kerr}, has been attributed to the fact that a \ac{kds} black hole has a larger event horizon compared to that of a \ac{rkds}.\\
When an observer is inclined away from the equatorial plane as illustrated in \cref{fig:difftheta}, the behaviour of the shadows is still consistent with that observed in \cref{fig:mineovale}. For high values of black hole spin, the shadows will have a near circular shape when the observer is inclined away from the equatorial plane. However, as the angle of inclination moves towards $\pi/2$, the left side of the shadow begin to appear more flattened as seen in \cref{fig:diffTheta}. Thus, a change in the angle of inclination will  change the quantitative aspect of the shadows i.e, the horizontal and vertical diameters of the shadows will change with the change in angle of inclination. The analytic behaviour with respect to changes in $a$ and $\Lambda$ will remain the same. 
\begin{figure}[htb]
	\centering 
	\begin{subfigure}{0.5\textwidth}
		\includegraphics[width=0.7\linewidth]{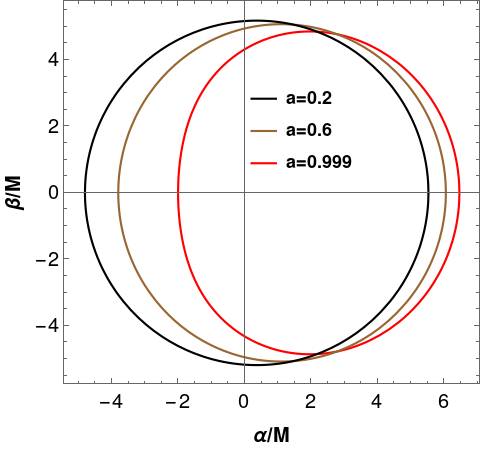}
		\caption{\ac{kds} Black Hole shadow for celestial coordinates \cref{50} and (\ref{51}) when the observer's angle of inclination is $\pi/3$ }
		\label{fig:kds3}
	\end{subfigure}\hfil
	\begin{subfigure}{0.5\textwidth}
		\includegraphics[width=0.7\linewidth]{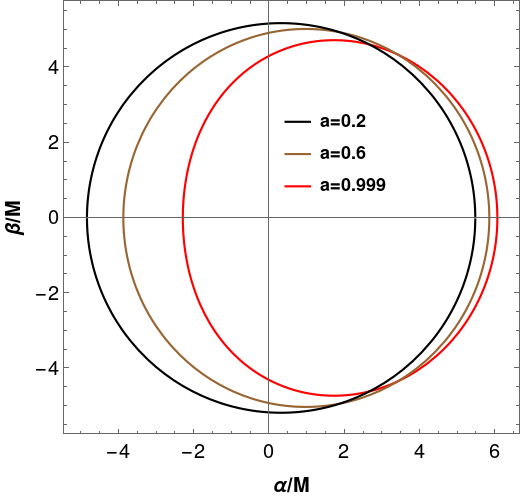}
		\caption{\ac{rkds} Black Hole shadow for celestial coordinates \cref{93} and (\ref{94}) when the observer is inclined at $\pi/3$}
		\label{fig:rkds3}
	\end{subfigure}\hfil
	\begin{subfigure}{0.5\textwidth}
		\includegraphics[width=0.7\linewidth]{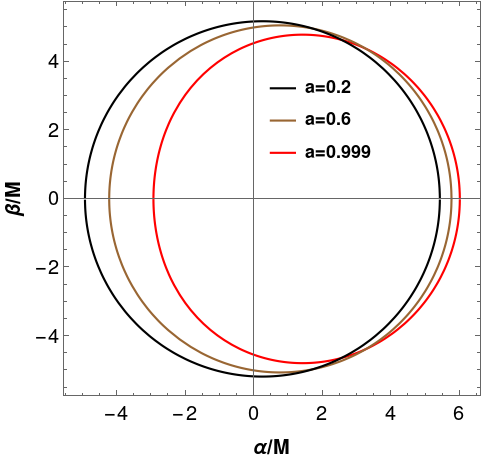}
		\caption{\ac{kds} Black Hole shadow for celestial coordinates \cref{50} and (\ref{51}) when the observer's angle of inclination is $\pi/5$ }
		\label{fig:kds5}
	\end{subfigure}\hfil
	\begin{subfigure}{0.5\textwidth}
		\includegraphics[width=0.7\linewidth]{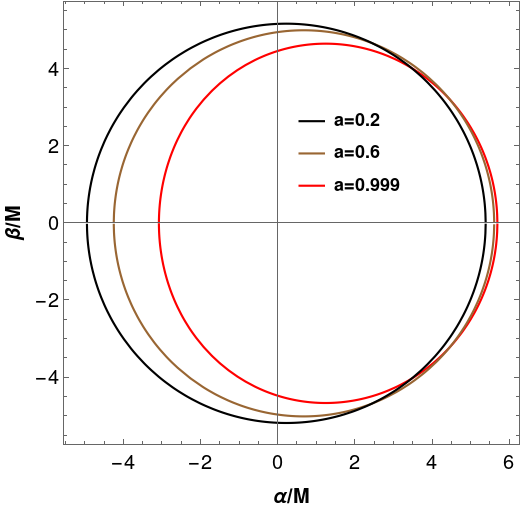}
		\caption{\ac{rkds} Black Hole shadow for celestial coordinates \cref{93} and (\ref{94}) when the observer is inclined at $\pi/5$}
		\label{fig:rkds5}
	\end{subfigure}\hfil
	\caption{Plots for different values of black hole spin when the observer's inclination is away from the equatorial plane. $M=1.06$  and $\Lambda=0.06m^{-2}$.
	}
	\label{fig:difftheta}
\end{figure}
\begin{figure}[htb]
	\centering 
	\begin{subfigure}{0.35\textwidth}
		\includegraphics[width=1.0\linewidth]{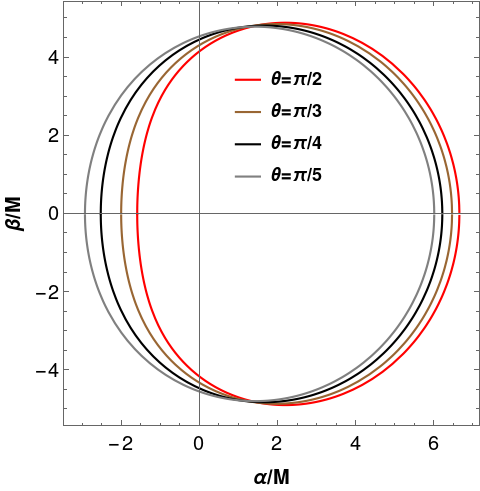}
		\caption{\ac{kds} Black Hole shadows for celestial coordinates \cref{50} and (\ref{51}).}
		\label{fig:kdst}
	\end{subfigure}\hfil
	\begin{subfigure}{0.35\textwidth}
		\includegraphics[width=1.0\linewidth]{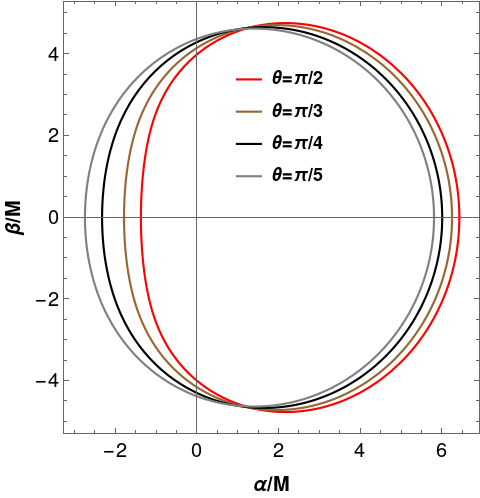}
		\caption{\ac{rkds} Black Hole shadows for celestial coordinates \cref{93} and (\ref{94})}
		\label{fig:rkdst}
	\end{subfigure}\hfil
	\caption{Plots  of a \ac{kds} with \ac{rkds} black hole shadows with $a=0.999$, $M=1.06$ and $\Lambda=0.06m^{-2}$ for different angles inclination.
	}
	\label{fig:diffTheta}
\end{figure}
\section{Intrinsic Curvature of the Shadows}
The black hole shadow is a one dimensional closed curve parametrized by $(\alpha(r),\beta(r))$. Making use of differential geometry concepts of closed curves, the intrinsic properties of the black hole shadow can be obtained. In this section, we will employ differential geometry to determine the curvature radius of the shadows. The concept of curvature radius was introduced in Ref. \cite{wei2019intrinsic} together with its applications in constraining the black hole's parameters. Moreover, by utilizing the symmetry of a black hole shadow, Ref. \cite{wei2019curvature} have obtained characteristic points along the one dimensional curve of the shadow. Using the characteristic points and curvature radius, the Kerr black hole parameters have been obtained. An illustration of the characteristic points is as shown in \cref{fig:characpoints}. We will obtain the curvature radius and characteristic points of a \ac{kds} and \ac{rkds} black hole shadow. Using these properties, we will do a quantitative comparison of these shadows. We will further compare this quantitative behaviour to that of a Kerr black hole  \cite{wei2019curvature}.

 Utilizing the concept of parametric curves, the radius of curvature can be obtained by,
\begin{align}
R_{curvature}= \left| \dfrac{(\alpha'(r)^{2}+\beta'(r)^2)^{3/2}}{\alpha'(r)\beta''(r)-\beta'(r)\alpha''(r)} \right|. \label{radcurv}
\end{align}
We then compute the curvature radius at points $T$,$D$,$R$ and $B$ as shown in \cref{fig:characpoints} and denote them as $R_{T}$, $R_{D}$, $R_{R}$, $R_{B}$ respectively. To avoid writing the cumbersome equations that we obtain, we give the general approach by which we calculate the points,
\begin{enumerate}
    \item 
    To measure the curvature radius at point $T$ and $B$, we work with a value of $r$ obtained through $\partial_{r}\beta=0$.
    \item In evaluating curvature radius at $D$ and $R$, we use $r$ for which $\beta=0$.
    \item The vertical diameter, $\Delta \beta=2\beta_{T}$.
    \item Horizontal diameter, $\Delta \alpha = \sqrt{(\alpha_{D}-\alpha_{R})^2}$.
\end{enumerate}
We have numerically computed these points for a \ac{kds} and \ac{rkds} black hole shadow. Additionally, with the aid of the value of gravitational radius obtained from \ac{eht} observations \cite{event2019firsta},\cite{wei2019curvature},
\begin{align}
\theta_{g}= \frac{G M}{c^{2}D}=3.8 \mu a s, \label{eq:c}
\end{align}
we model M87* to these shadows.
 The results are in \cref{table:Lm52} and (\ref{table:L01}) for two different values of $\Lambda$.
 
\begin{table}[ht]
	\caption{Points evaluated for  $\theta=16^\circ$, $a=0.5$ and $\Lambda=1.11\times 10^{-52}m^{-2}$} .
	\centering 
	\begin{tabular}{c c c c c c c} 
		\hline\hline 
		points& Kerr & \ac{kds}, ($\ref{50}$,$\ref{51}$) & \ac{rkds}, ($\ref{93}$,$\ref{94}$) & \ac{kds} \cite{ovalle2021kerr},($\ref{204}$,$\ref{205}$)&  \ac{rkds},\cite{ovalle2021kerr},($\ref{204}$,$\ref{205}$)\\ [0.5ex] 
		\hline 
		$\Delta \beta(\mu a s) $  &38.9617  & 38.9617 &38.9617  &38.9617  & 38.9617   \\ 
		$\Delta \alpha(\mu a s) $  &38.9123  & 38.9123 &38.9123  &38.9123  & 38.9123   \\
		$R_{T} (\mu a s) $  &19.4318  & 19.4318 &19.4318  &19.4318  & 19.4318   \\
		$R_{D} (\mu a s) $  &19.5096  & 19.5099 &19.5099  &19.5099  & 19.5099   \\
		$R_{R} (\mu a s) $  &19.502  & 19.502 &19.502  &19.502  & 19.502   \\
	\end{tabular}
	\caption*{We have chosen $\Lambda=1.11\times 10^{-52}m^{-2}$ because cosmological tests and  CMB measurements imply that this is the  relevant value of the cosmological constant \cite{spergel2003first}, \cite{aghanim2020planck},\cite{stepanian2021black}}
	\label{table:Lm52} 
\end{table}
\begin{table}[ht]
	\caption{Points evaluated for  $\theta=16^\circ$, $a=0.5$ and $\Lambda=0.06m^{-2}$} 
	\centering 
	\begin{tabular}{c c c c c c c} 
		\hline\hline 
		points&  \ac{kds}, ($\ref{50}$,$\ref{51}$) & \ac{rkds}, ($\ref{93}$,$\ref{94}$) & \ac{kds} \cite{ovalle2021kerr},($\ref{204}$,$\ref{205}$)&  \ac{rkds},\cite{ovalle2021kerr},($\ref{204}$,$\ref{205}$)\\ [0.5ex] 
		\hline 
		$\Delta \beta (\mu a s) $   & 38.6587 &38.3582 &55.1325  & 54.963   \\ 
		$\Delta \alpha (\mu a s) $   & 38.6533 &38.267 &56.2004  & 55.0885   \\
		$R_{T} (\mu a s) $   & 19.2878 &19.0882 &28.1495  & 27.6076   \\
		$R_{D} (\mu a s) $   & 19.4078 &19.2278 &27.9915  & 27.4234   \\
		$R_{R} (\mu a s) $   & 19.4021 &19.2221 &27.9833  & 27.4154   \\
	\end{tabular}
	\label{table:L01}  
\end{table}

In \cref{table:Lm52},  the second column contains values that we obtained by modelling M87* to a Kerr black hole shadow. The values agree with those obtained by \cite{wei2019curvature}. We however note that, in \cite{wei2019curvature}, the values of $R_{T}$ and $R_{R}$ given in table 2 seem to have been interchanged. In the third and fourth column, we have the \ac{kds} and \ac{rkds} values using the celestial coordinates obtained in our calculation, \cref{50}-(\ref{51}) and \cref{93}-(\ref{94}) respectively. The fifth and sixth column has values that we calculated using the celestial coordinates used in Ref. \cite{ovalle2021kerr}, \cref{204}-(\ref{205}). The value of the cosmological constant we have used on this table (for \ac{kds} and \ac{rkds}) is $\Lambda=1.11\times 10^{-52}m^{-2}$. We observe that for this value of the cosmological constant, the values of curvature radius at points $D$, $R$ and $T$ are approximately equal, across the table. The same applies to values of the horizontal and vertical diameters. Thus, for $\Lambda=1.11\times 10^{-52}m^{-2}$, the \ac{kds} and \ac{rkds} black hole shadow is indistinguishable from that of a Kerr black hole shadow. 

For \cref{table:L01}, we have used a larger value of the cosmological constant, $\Lambda=0.06m^{-2}$. The second and third column represents values calculated using the celestial coordinates we obtained in our work, \cref{50}-(\ref{51}) and \cref{93}-(\ref{94}). The fourth and fifth column on the other hand are values we have computed using celestial coordinates in Ref.\cite{ovalle2021kerr}, \cref{204}-(\ref{205}). The values are no longer equal across the table and the discrepancy is now obvious.
 Comparing the second and third column with the corresponding column in \cref{table:Lm52} (third and forth column), we observe that the radii of curvature and diameters decrease when a larger value of $\Lambda$ is used. Thus, in both \ac{kds} and \ac{rkds} space-times the shadow will decrease in size for a larger value of $\Lambda$, in accordance with the celestial coordinates we obtained, \cref{50}-(\ref{51}) and \cref{93}-(\ref{94}). By using orthonormal tetrads approach, Ref. \cite{li2020shadow} results also show that the size of a \ac{kds} shadow would decrease when $\Lambda$ is increased.\\
 However, comparing the forth and fifth column with the corresponding column in \cref{table:Lm52} (fifth and sixth column), the radii of curvature and diameters increase with increase in the value of $\Lambda$.\\
 From \cref{table:L01}, the values of curvature radii, horizontal and vertical diameters of a \ac{kds} black hole shadow are larger than those of a \ac{rkds} shadow despite the difference being small. This is in agreement with \cref{fig:dscomparison} where we observed that a \ac{kds} black hole casts a larger shadow than  a \ac{rkds} black hole. Thus, the \ac{rkds} solution shows a gravitational field less intense than the standard \ac{kds} solution.\\
It is worth to note that besides the approach of intrinsic curvature, there exist other alternative ways of comparing black hole shadows as illustrated in Ref. \cite{junior2021can} and \cite{mars2017fingerprints}.
\section{Constraints on Kerr-de Sitter and Kerr-de Sitter Revisited black hole.}
Recently, in Ref. \cite{kocherlakota2021constraints}, the 2017 \ac{eht} observations of M87* were utilized and a constraint on the characteristic radius of the shadow was obtained at $68\%$ confidence level. It was shown that irrespective of the underlying solution being spherically or axially symmetric, the radius of the shadow ought to lie in the range,
\begin{align}
    4.31M \approx r_{sh, EHT-min} \leq \tilde{r}_{sh}, \quad r_{sh,A} \leq r_{sh,EHT-max} \approx 6.08M, \label{constraint}
\end{align}
where $\tilde{r}_{sh}$ and $r_{sh,A}$ denotes radius of the shadow in spherically symmetric and axially symmetric solution respectively. From this shadow size, highly charged dilaton black holes were ruled out of M87* observations. Additionally, in Ref.\cite{zakharov2021constraints}  this observation was used to constrain the tidal charge of a  Reissner–Nordstrom black hole. \\
In this section we will implement the concept of curvature radius that we have discussed in the previous section to this observation. In so doing we will constrain the parameters of a \ac{kds} and \ac{rkds} black hole. In  Ref. \cite{wei2021constraining},  the concept of curvature radius has been used to constrain a Kerr and a Kerr-Newman black hole. 

 \begin{figure}[htb]
	\centering 
	\begin{subfigure}{0.5\textwidth}
		\includegraphics[width=0.8\linewidth]{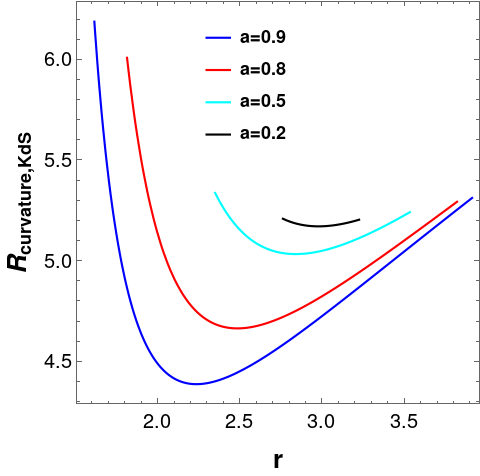}
		\caption{ Kerr-de Sitter black hole shadow radius of \\curvature for fixed $\theta=\pi/2$ }
		\label{fig:RKDtp2}
	\end{subfigure}\hfil
	\begin{subfigure}{0.5\textwidth}
		\includegraphics[width=0.8\linewidth]{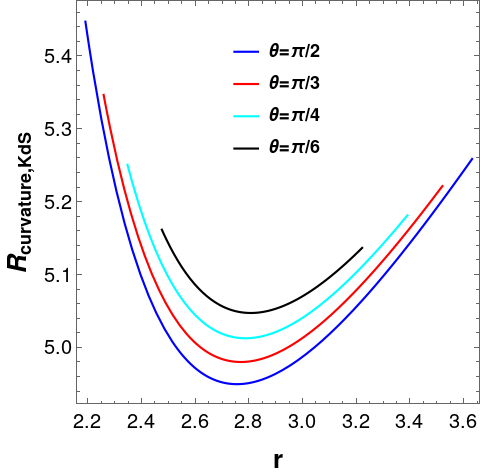}
		\caption{ Kerr-de Sitter black hole shadow radius of \\curvature for fixed spin, $a=0.6$}
		\label{fig:RKDa06}
	\end{subfigure}\hfil
	\begin{subfigure}{0.5\textwidth}
		\includegraphics[width=0.8\linewidth]{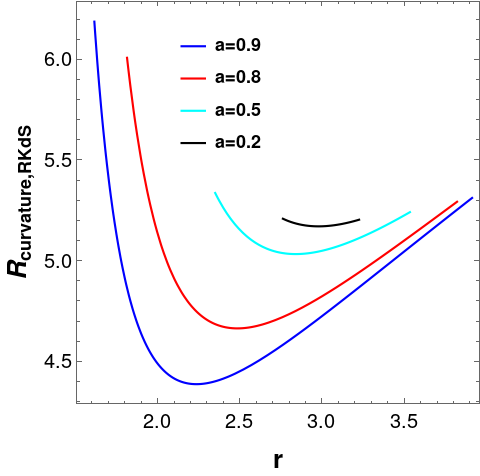}
		\caption{Kerr-de Sitter Revisited black hole shadow radius of\\ curvature for fixed $\theta=\pi/2$}
		\label{fig:delarkdlpm52}
	\end{subfigure}\hfil
	\begin{subfigure}{0.5\textwidth}
		\includegraphics[width=0.8\linewidth]{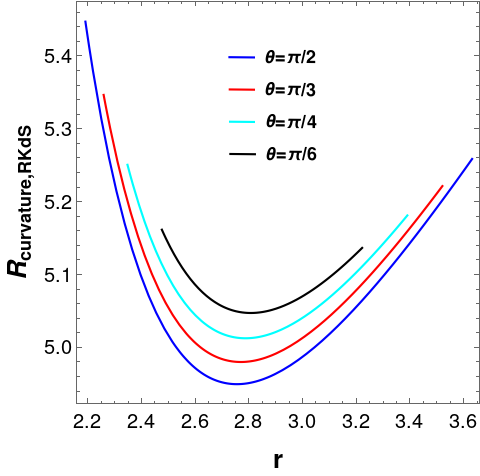}
		\caption{ Kerr-de Sitter black hole shadow radius of \\curvature for fixed spin, $a=0.6$.}
		\label{fig:RrKDa06}
	\end{subfigure}\hfil
	\caption{Radius of curvature for Kerr-de Sitter and Kerr-de Sitter Revisited black hole shadow as a function of $r$. $\Lambda=1.11\times 10^{-52} m^{-2}$and $r \in [r_{A},r_{B}]$ is such that $\partial_{r} \beta=0$. }
	\label{fig:R}
\end{figure}
While excluding $Z_2$ symmetry, the radius of curvature will have one maximum and one minimum \cite{wei2019intrinsic}. \autoref{fig:R} illustrates the maximum and minimum points of the curvature radius of a \ac{kds} and \ac{rkds} black hole shadow. From this figure, the radius of curvature has a
local maximum at $r=r_A$ and $r=r_B$. Furthermore, as the black hole spin and $\theta$ increases, these two local maxima also increase. The local maximum at $r=r_{A}$ is greater than that at $r=r_{B}$. Hence $R_{max}=R(r_{A})$. Additionally, the minimum of the radius of curvature ($R_{min}$) forms at the well of these curves. This minimum point decreases with increase in spin and $\theta$. $R_{min}$ and $R_{max}$ then gives a lower and an upper bound on the size of the shadow.  Thus, \cref{constraint} can be expressed as \cite{wei2021constraining},
\begin{align}
    4.31M \leq R_{min}, \quad R_{max} \leq 6.08M.\label{realconstraint}
\end{align}
In other words, $R_{min}$ should not decrease below $4.31M$ and $R_{max}$ should not increase beyond $6.31M$. If this happens, then such values should be excluded from observations of M87*.
Using \cref{realconstraint}, we obtain regions that are excluded from observations thus providing constraints on the black hole parameters.\\
\begin{figure}[htb]
	\centering 
	\begin{subfigure}{0.5\textwidth}
		\includegraphics[width=0.8\linewidth]{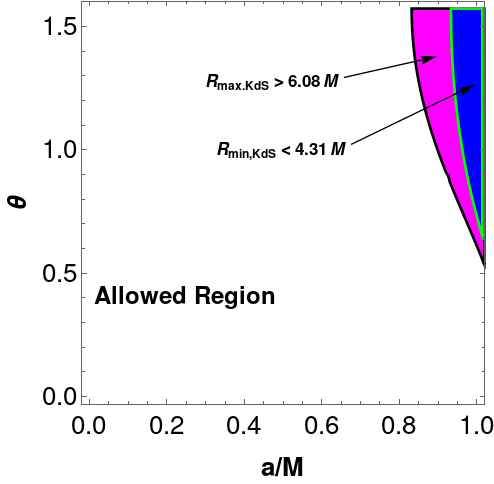}
		\caption{ Excluded and permitted regions for a shadow cast by a Kerr-de Sitter black hole. }
		\label{fig:RCKD}
	\end{subfigure}\hfil
	\begin{subfigure}{0.5\textwidth}
		\includegraphics[width=0.8\linewidth]{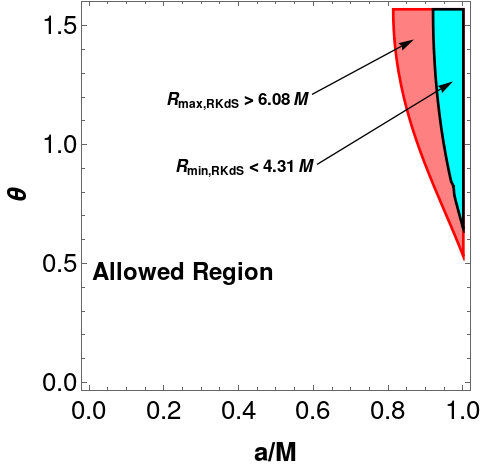}
		\caption{ Excluded and permitted regions for a shadow cast by a Kerr-de Sitter Revisited black hole.}
		\label{fig:RCrKD}
	\end{subfigure}\hfil
	\caption{Region Plots for the radius of curvature, $\Lambda=1.11\times10^{-52}m^{-2}$. The shaded parts indicate regions where the minimum radius of curvature is less than $4.31M$ and the maximum curvature is greater than $6.08M$. Such regions according to \cref{realconstraint} are excluded from observations by M87* hence imposing constraints on $\theta$ and $a$. The unshaded part is the allowed region.}
	\label{fig:Rconstraints}
\end{figure}
From \cref{fig:Rconstraints}, we observe that in both black holes, excluded regions appear at high black hole spin $a/M>0.812311$ and larger angles of inclination $\theta>0.532512 \approx 30.5107 ^{\circ}$. However, for small angles of inclination, no excluded regions occur. Thus, for a \ac{kds} and \ac{rkds} black hole, small angles of inclination pass the constraints of M87* observations. On the other hand, when $a/M>0.812311$, large angles of inclination do not pass the constraints.

 Practically, there are no sensible differences in constraints between $\Lambda=0$ (\cite{bambi2019testing},\cite{wei2021constraining}) and $\Lambda \sim 10^{-52}$ , this section thus confirms the findings of \cite{stepanian2021black}.

\clearpage
\newpage
 \section{Conclusion}
We have analyzed black hole shadows in \ac{kds} and \ac{rkds} space-times for observers located in the vicinity of the static radius. Embedding diagrams of both ordinary geometry and optical reference geometry have shown that the space-time in the proximity of the static radius in de Sitter space-time is analogous to an asymptotically flat space-time. This makes such observers to be of great interest to our work.
In view of the fact that a black hole shadow forms as a result of the unstable photon region being projected on the observer's sky, we have first investigated the behaviour of photon orbits in the respective photon regions.

Particularly, we have investigated the behaviour of the radius of equatorial circular photon orbits which form the lower and upper bound of the photon region. In both space-times, the radius of equatorial prograde (retrograde) circular photon orbits decreases (increases) with increase in black hole spin. For increasing value of $\Lambda$, the radius of equatorial prograde (retrograde) circular photon orbits increases (decreases). We have done a qualitative analysis of the shadows with respect to this behaviour. We find that the shadows reflect this observed behaviour of the radius of equatorial circular photon orbits. Besides, we have compared our result with Ref.\cite{ovalle2021kerr}. For varying black hole spin, our results agree qualitatively. However, for varying $\Lambda$, the impact parameters considered in Ref.\cite{ovalle2021kerr} yield shadows whose behaviour contradicts the behaviour of the radius of equatorial circular photon orbits.

Furthermore, we have numerically computed the radii of curvature at specific characteristic points along the curve of a \ac{kds} and \ac{rkds} black hole shadow. We have further obtained numerical values for the horizontal and vertical diameters of the shadows. These values have then been modelled to M87* observations. For $\Lambda=1.11\times 10^{-52} m^{-2}$, we observe that  \ac{kds} and \ac{rkds} black hole shadows have  radii of curvature, horizontal and vertical diameters  that are approximately equal. Further, these values are   approximately equal to those of a Kerr black hole shadow. Thus, astrophysically relevant observations (observations for which $\Lambda=1.11 \times 10^{-52} m^{-2}$) cannot distinguish between a \ac{rkds}, \ac{kds} and Kerr black hole shadow. For $\Lambda=0.06 m^{-2}$, see \cref{table:L01}, the quantitative discrepancy between our result and Ref. \cite{ovalle2021kerr} becomes obvious. But such immense value for $\Lambda$ is not realistic. The values also indicate that a \ac{rkds} black hole casts a smaller shadow than that of a \ac{kds} black hole despite the difference being small. Generally, considering our results, an increase in the cosmological constant decreases the size of a \ac{kds} and \ac{rkds} black hole shadow. 

Finally, utilizing the constraint on the characteristic areal radius of the shadow obtained by the \ac{eht} collaboration, we have constrained a \ac{kds} and \ac{rkds} black hole. We find that, for $a/M>0.812311$, large angles of inclination $\theta>30.5107^{\circ}$ are excluded from M87* observations in both \ac{kds} and \ac{rkds} black holes.

\clearpage
\newpage
\begin{acknowledgments}

The authors thank FAPES/CNPq/CAPES and Proppi/UFOP for financial support. The authors would also like to thank Z. Stuchlik and J. Ovalle for their comments.
\end{acknowledgments}

\bibliography{Refs}

\end{document}